\documentclass{emulateapj}
\pdfoutput=1

\usepackage{graphicx}
\usepackage{amsmath}
\usepackage{amssymb}
\usepackage{fancyvrb}
\usepackage{epsf}
\usepackage{sidecap}
\usepackage{natbib}
\usepackage{aas_macros}
\usepackage[bookmarks,bookmarksnumbered,colorlinks=true, citecolor=blue, linkcolor=black]{hyperref}

\newcommand{\hmpc}{{\,h^{-1}\rm\,Mpc\,}}
\newcommand{\hkpc}{{\,h^{-1}\rm\,kpc\,}}
\newcommand{\hmpcC}{{\,h^{3}\rm\,Mpc^{-3}\,}}
\newcommand{\etal}{{et al.\,}}

\shorttitle{PRIMUS: The Effect of Physical Scale on the Luminosity-Dependence of Galaxy Clustering}
\shortauthors{Bray et al.}

\begin{document}

\title{PRIMUS: The Effect of Physical Scale on the Luminosity-Dependence of Galaxy Clustering via Cross-Correlations}

\author{
Aaron~D.~Bray\altaffilmark{1}, 
Daniel~J.~Eisenstein\altaffilmark{1}, 
Ramin~A.~Skibba\altaffilmark{2},
Michael~R.~Blanton\altaffilmark{3}, 
Alison~L.~Coil \altaffilmark{2}, 
Richard~J.~Cool\altaffilmark{4}, 
Alexander~J.~Mendez\altaffilmark{5}, 
John~Moustakas\altaffilmark{6},  
Guangtun~Zhu\altaffilmark{5,}\altaffilmark{7}
}

\altaffiltext{1}{Harvard-Smithsonian Center for Astrophysics, 60 Garden St., Cambridge, MA 02138, USA ; abray@cfa.harvard.edu}
\altaffiltext{2}{Department of Physics, Center for Astrophysics and Space Sciences, University of California, 9500 Gilman Dr., La Jolla, San Diego, CA 92093, USA}
\altaffiltext{3}{Center for Cosmology and Particle Physics, Department of Physics, New York University, 4 Washington Place, New York, NY 10003, USA}
\altaffiltext{4}{MMT Observatory, 1540 E Second Street, University of Arizona, Tucson, AZ 85721, USA}
\altaffiltext{5}{Department of Physics $\&$ Astronomy, Johns Hopkins University, 3400 N. Charles Street, Baltimore, MD 21218, USA}
\altaffiltext{6}{Department of Physics and Astronomy, Siena College, 515 Loudon Road, Loudonville, NY 12211, USA}
\altaffiltext{7}{Hubble Fellow}

\begin{abstract}
We report small-scale clustering measurements from the PRIsm MUlti-object Survey (PRIMUS) spectroscopic redshift survey as a function of color and luminosity. We measure the real-space cross-correlations between $62$,$106$ primary galaxies with PRIMUS redshifts and a tracer population of $\sim 545$,$000$ photometric galaxies over redshifts from $z = 0.2$ to $z = 1$. We separately fit a power-law model in redshift and luminosity to each of three independent color-selected samples of galaxies. We report clustering amplitudes at fiducial values of $z = 0.5$ and $L = 1.5 L^{*}$. The clustering of the red galaxies is $\sim\!3$ times as strong as that of the blue galaxies and $\sim\!1.5$ as strong as that of the green galaxies. We also find that the luminosity dependence of the clustering is strongly dependent on physical scale, with greater luminosity dependence being found between $r = 0.0625 \hmpc$ and  $r = 0.25 \hmpc$, compared to the $r = 0.5 \hmpc$ to  $r = 2 \hmpc$ range. Moreover, over a range of two orders of magnitude in luminosity, a single power-law fit to the luminosity dependence is not sufficient to explain the increase in clustering at both the bright and faint ends at the smaller scales. We argue that luminosity-dependent clustering at small scales is a necessary component of galaxy-halo occupation models for blue, star-forming galaxies as well as for red, quenched galaxies.

\end{abstract}


\section{Introduction}

Because baryons are a subdominant component of the matter density of the universe in $\Lambda$CDM cosmology, galaxies form within dark matter overdensities. Precise measurements of galaxy clustering can thus be used to probe the underlying dark matter structure, and the dependence of clustering on galaxy properties can be used to examine the connection between galaxy formation and the large scale structure environment.

Redshift surveys supply increasingly plentiful data, and galaxy clustering measurements continue to offer one of the best ways to interpret these surveys in the context of the $\Lambda$CDM framework. At low redshifts, for example, the 2-degree Field Galaxy Redshift Survey (\citealt{Colless01}), the Sloan Digital Sky Survey (SDSS; \citealt{York00}), and the 6-degree Field Galaxy Survey (\citealt{Jones09}) measured hundreds of thousands of redshifts over tens of thousands of square degrees out to $z \sim 0.2$. At intermediate redshifts, the wide-field AGN and Galaxy Evolution Survey (AGES; \citealt{Kochanek12}) measured tens of thousands of redshifts over almost eight square degrees, and at higher redshifts, surveys such as the DEEP2 Galaxy Redshift Survey (DEEP2; \citealt{Newman13}), zCOSMOS (\citealt{Lilly07}), and the VIMOS-VLT (Very Large Telescope) Deep Survey (\citealt{LeFevre05}) have measured tens of thousands of redshifts each over fields up to a few square degrees.

Previous studies have established that the galaxy clustering signal depends on observational quantities such as morphology, luminosity, color, and their physical analogs such as stellar mass and star formation rate. For example, \citet{DG76} measured steeper autocorrelation functions for elliptical than spiral galaxies, \citet{Dressler80} provided evidence that galaxies with more luminous spheroidal components preferred higher density regions, and \citet{White88} showed that, in agreement with Cold Dark Matter models, galaxies with higher circular velocities traced high density environments. More recently, observations from the local universe out to redshifts of $z \sim 1$ demonstrate that red, passive galaxies form are more highly clustered than blue, star-forming galaxies (e.g., \citealt{Zehavi05}; \citealt{Coil08}; \citealt{Skibba09}), and that galaxies with higher luminosity and stellar mass are more clustered than those with lower luminosity and mass (e.g., \citealt{Norberg01}; \citealt{Coil06}; \citealt{Meneux06}; \citealt{Meneux09}; \citealt{Coupon12}; \citealt{Marulli13}). In particular, the faint and bright ends of the luminosity spectrum of red galaxies show increased clustering (e.g., \citealt{Hogg03}; \citealt{Eisenstein05}; \citealt{Swanson08a}; \citealt{Zehavi11}). Using photometric redshift surveys, the dichotomy between quiescent and star-forming galaxies, and the emergence of their differential clustering, has been studied out to beyond redshift of $z = 3$ (\citealt{Williams09}; \citealt{Hartley10}; \citealt{Hartley13}).

Into this context, the PRIsm Multi-object Survey (PRIMUS; \citealt{Coil11}; \citealt{Cool13}) measures $\sim\!130$,$000$ redshifts from $z = 0.2$ out to $z = 1.2$ over almost ten square degrees. With a larger survey area to reduce cosmic variance and more depth than previous intermediate redshift surveys, PRIMUS allows for the measurement of the evolution of galaxy properties over this redshift range and of clustering as a function of galaxy properties out to $z \sim 1$. Only the VIMOS Public Extragalactic Redshift Survey (VIPERS; \citealt{Guzzo14}) is comparable in its targeting of a similar number of redshifts out past $z \sim 1$ over a wide survey area, although to a slightly lower depth. 

The first clustering results from PRIMUS \citep{Skibba14} used auto-correlations to measure the galaxy clustering as a function of luminosity and color over projected scales of $0.1 \hmpc \leq r \leq 20 \hmpc$. This work extends those results to smaller scales and examines the luminosity dependence as a function of color and scale. We employ the cross-correlation methodology of Eisenstein (2003; hereafter E03) to measure the real-space, deprojected clustering of $62$,$106$ PRIMUS galaxies with respect to $\sim L^*$ tracer galaxies drawn from a parent population of $\sim 545$,$000$ photometric galaxies from the imaging catalogs that overlaps the PRIMUS footprint. Eisenstein \etal (2005; hereafter E05) previously applied this technique to LRGs in SDSS; we now extend this to a wider range of luminosities and colors, to smaller physical scales, and to higher redshifts.

By cross-correlating our primary galaxies against a sample nearly ten times larger, we avoid the Poisson noise inherent in the autocorrelation of small subsamples of galaxies. For example, it is less important, in our case, that bins in luminosity be of relatively equal size, because the secondary sample defines the environmental densities around each population in precisely the same manner, and so even bins with relatively small numbers of primary galaxies have small error bars. This is important because we wish to measure the clustering out to $z = 1$, where we must contend with small total numbers of galaxies, while still measuring trends over subsamples in color and luminosity.

Precise measurements of small-scale clustering are crucial to understanding galaxy formation. These cross-correlation measurements bear directly on the connection between dark matter halo properties and galaxy properties predicted and interpreted with analytic and semi-empirical models of galaxy formation (e.g., \citealt{PS00}; \citealt{BW02}; \citealt{Yang03}; \citealt{Zehavi05}; \citealt{Conroy06}; \citealt{SS09}; \citealt{Behroozi10}; \citealt{Moster10}; \citealt{HW13}; \citealt{Masaki13}), by probing the ``one-halo'' term and the transition to the ``two-halo" term. By examining jointly the relative clustering of color and luminosity selected galaxy samples over these scales, we can study how galaxy evolution depends on the local environment. In this paper (Paper I), we present our initial results along with interpretations in the context of halo models. In a follow-up paper (Paper II), we will address the presence of galactic conformity (see \citealt{Weinmann06}; \citealt{Kauffmann13}; \citealt{Hearin14}; \citealt{Hartley15}) in the PRIMUS sample.

The paper proceeds as follows: Section 2 provides a detailed explanation of the cross-correlation methodology. Section 3 overviews the PRIMUS survey and presents the sample selection for both the spectroscopic and imaging samples. We present our results in Section 4 and interpret them and compare them to the literature in Section 5. We conclude in Section 6 by highlighting our main results and suggesting avenues for future investigation. Throughout this paper, we use an $\Omega_m = 0.3$ flat cosmology with $h = 1$. We use AB magnitudes with dust reddening corrections applied \citep{SFD98}.

\section{Cross-correlation Methodology}
We measure the real-space galaxy clustering using the angular cross-correlation methodology described in E03. This requires both a primary galaxy sample with known redshifts and a significantly larger catalog of photometric galaxies. In our case, the \textit{primary} sample uses PRIMUS galaxies with spectroscopic redshifts (Section 3.4), and the \textit{secondary} sample uses galaxies from the imaging catalogs that overlap the PRIMUS field. The imaging catalog provides a tracer population with which to measure the real-space density around each spectroscopic galaxy. We calculate the cross-correlation signal for each \textit{primary} by assuming that all imaging galaxies are at the same redshift as the spectroscopic galaxy. This allows us to calculate passively evolved, k-corrected absolute magnitudes (see Section 3.3) and projected distances for each imaging galaxy. We then cross-correlate with the spectroscopic PRIMUS primary galaxy only those imaging galaxies in a fixed luminosity bin, so as as to have a tracer population with a well-understood number density. Because foreground and background galaxies in the secondary sample have no physical correlation with the spectroscopic galaxy, only those imaging galaxy that are at the same redshift will contribute to the clustering signal; E05 shows gravitational lensing can be neglected.

\begin{figure*}[t]
\plotone{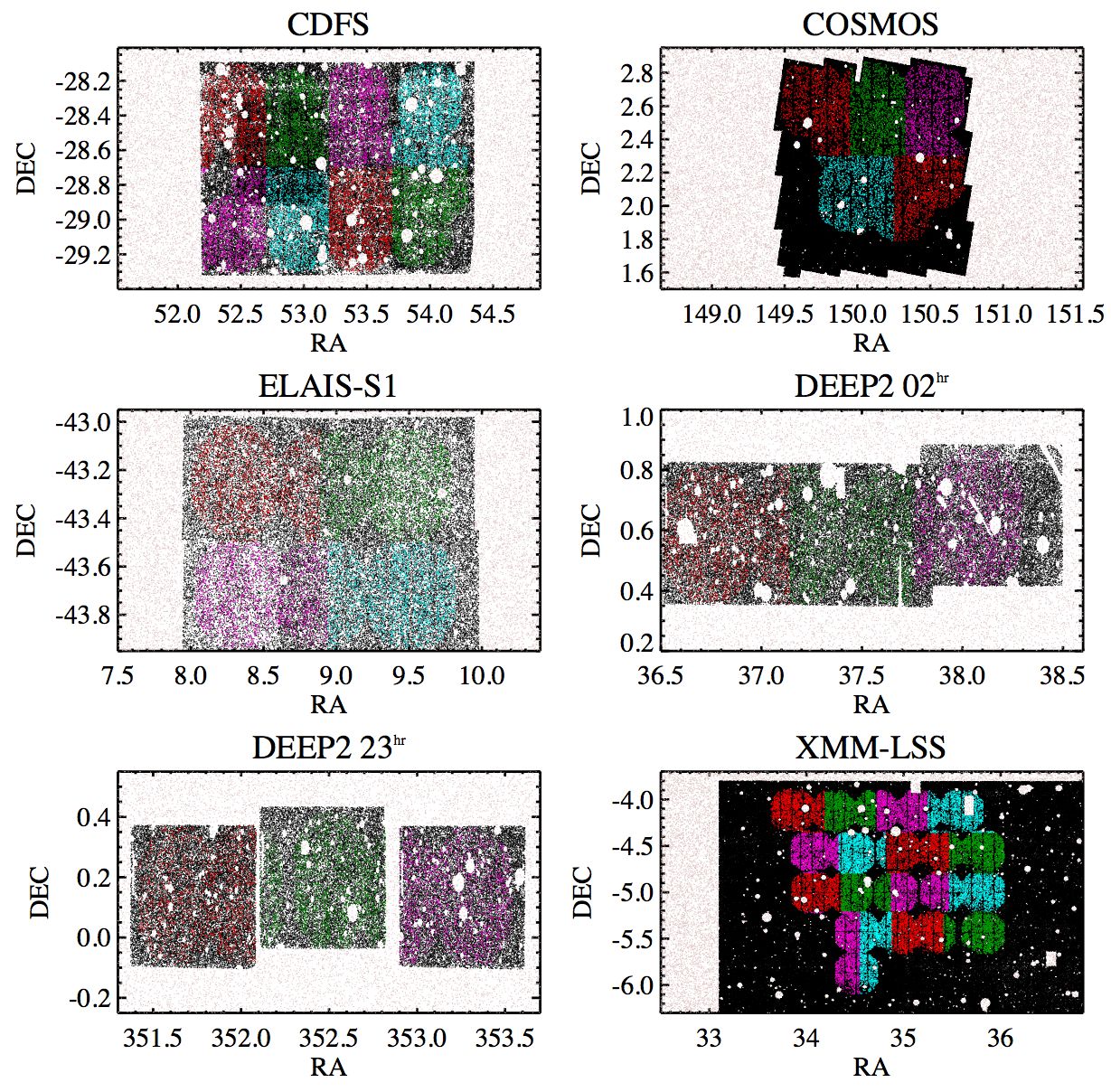}
\caption{\label{fig:fig1}
\small 
PRIMUS coverage from the six science fields used in this work. Galaxies are shown in color; red, green, magenta, and cyan are used only to distinguish between the spatial jackknife regions used for error estimation, and so repeated colors are not representative of any physical difference. Each jackknife region has approximately $1$,$500$ primary spectroscopic PRIMUS galaxies. Black points show the secondary imaging sample (sparse sampled by 50$\%$ for visual clarity.) Tan points are randoms distributed in the masked regions. At most, $100$,$000$ random points are shown, but usually many fewer. The masked border regions extends beyond the edges of the figures, but are not shown for clarity. In fact, the masked regions are more densely sampled than photometric regions.
}
\end{figure*}

\subsection{Computational Details}
By assuming spherical isotropy, angular clustering measurements can be inverted into a measure of the real-space clustering, $\xi(r)$, using an Abel transformation \citep{vonZeipel08}. This deprojection, however, requires computing the derivative of the angular clustering. Instead, the E03 method eliminates the need to calculate this noisy derivative by integrating the correlation function over a spherical three-dimensional window:
\begin{equation}
\Delta(a) = \frac{1}{V(a)}\int_0^\infty  \! 4 \pi r^2 \xi(r) W(r, a) \, \mathrm{d} r
\end{equation}
where the weighted volume $V(a) = \int_0^\infty  \! 4 \pi r^2 W(r, a) \, \mathrm{d} r$ and $W(r, a)$ is the smoothing window, such that the correlation function is weighted by $W$ at a distance $r$, given a scale length $a$.  $\Delta(a)$ is the overdensity of objects from the imaging catalog; to wit, a measurement of $\Delta(a) = 1$ would mean that the density of imaging objects in the window $W(r, a)$ was equal to the mean density of that imaging tracer population.

In this paper, we pick the same smoothing window as E05, which relative to a spherical Gaussian window, reduces the contribution of spectroscopic-imaging pairs at small angular distances to the weighted overdensity, $\Delta(a)$, thus reducing the contribution of systematics from photometric deblending.

\begin{equation}
W(r, a) = \frac{r^2}{a^2} \, \mathrm{exp} \, (-\frac{r^2}{2a^2})
\end{equation}

E03 shows that the $\Delta (a)$ statistic can be computed as a pairwise summation of the spectroscopic and the imaging catalogs, as:
\begin{eqnarray}
\Delta (a) &  =  & \frac{1}{N_{sp}}\sum\limits_{j\in\left\{sp\right\}} \, \frac{1}{\phi_{0}(z_{j})V} \sum\limits_{k\in\left\{img\right\}} \, G(R_{jk})  \\
            &  =  & \frac{1}{N_{sp}}\sum\limits_{j\in\left\{sp\right\}} \, \Delta_{j} (a),
\end{eqnarray}
where $N_{sp}$ is the number of primary spectroscopic objects, $\phi_{0} (z)$ is the real-space density of objects at redshift $z_j$, and $R_{jk}$ is transverse distance, and where the weighting function $G(R)$ is defined as:
\begin{eqnarray}
G(R) &  =  & \frac{1}{R}\frac{\mathrm{d}F}{\mathrm{d}R},  \\
F(R)  &  =  & \frac{2}{\pi}\int_0^R \frac{{r}^{2}W(R)}{\sqrt{R^2-r^2}} \mathrm{d}r,
\end{eqnarray}

This method allows us to compute a noisy estimate of the overdensity, $\Delta_{j} (a)$, around each and every spectroscopic object. These individual values can then be trivially averaged together over any subset of galaxies within the full sample. This permits quick measurements of the $\Delta (a)$ statistic with respect to many different dependent quantities --- such as luminosity, color, and redshift --- without the computational overhead involved in repeatedly calculating the angular clustering of different subsamples. In fact, without additional assumptions, the calculation returns $\phi_0 (z) \Delta (a)$, which is the overdensity multiplied by the number density of the tracer population. So in order to recover $\Delta (a)$, we need a model of the redshift evolution of the number density, which we define in Section 3.

To correct for excluded regions (bright stars, bad pixels, edges, etc.), we create a random galaxy catalog in which a dense set of galaxies are distributed within all survey gaps and around the borders of the imaging catalog masks. These random points only need be assigned an angular position on the sky (as opposed to a redshift, color, or luminosity). For each primary galaxy, $\Delta_{j} (a)$ is calculated for both galaxy-galaxy pairs and for galaxy-random pairs. These terms are summed together, but the galaxy-random sum carries an additional weight to account for the difference in projected densities between the imaging catalog and the random catalog. This is, in effect, a volume completeness correction, upweighting $\Delta_{j} (a)$ to account for any masked volume in the spherical window. 

We truncate the pair-wise summation at an inner radius of $5 \arcsec$, below photometric deblending is not sufficiently accurate.  We must also truncate the summation at an outer radius; for this, we choose $r = 9a$. This is large enough that the analytic correction to the sum is small, since $W(r, a)$ falls as a Gaussian in $r/a$; meanwhile, it is small enough that we can measure the cross-correlation function out to an effective radius of $2 \hmpc$ even at our lowest redshifts. To account for the inner and outer summation limits of our calculation, we must include in our computation an analytic correction term that is a function of $F(R)$. The full derivation and implementation are given in E03 and E05.

Lastly, to correct for completeness, each galaxy receives the PRIMUS \textit{primary} sample completeness weight. These weights include both \textit{a priori} magnitude- and spatial density-dependent sparse sampling, details of which can be found in \citet{Coil11} and \citet{Cool13}, as well as an \textit{a posteriori} spectroscopic success rate weight --- \textit{$f_{target}$} and \textit{$f_{collision}$}, respectively, as described in Equation 1 of \citet{Moustakas13}.

It is useful to be able to directly compare $\Delta (a)$ with $\xi(r)$. For an assumed power-law form of $\xi(r) \propto r^{\gamma}$, E05 note that the relationship between the two is given by:
\begin{equation}
\Delta(a) = \frac{2}{3}\sqrt{\frac{2}{\pi}}\sqrt{2}^{\,\gamma + 1}\Gamma\left(2 + \frac{\gamma + 1}{2}\right)\xi(a).
\end{equation}
In this work, we assume that $\gamma = -2$, so $\Delta(a) = \xi(a)/3 \approx \xi (1.73a)$. For ease of comparison with the literature, we thus define:
\begin{equation}
\xi_{_{\Delta}}(r) = \Delta(r/1.73)
\end{equation}
We report all scale-dependent measurements as a function of $r$. Thus, in order to obtain our results from $0.0625 \hmpc \leq r \leq 2.0 \hmpc$, we choose scale factors that range roughly from $0.036 \hmpc \leq a \leq 1.16 \hmpc$. Our statistic, $\xi_{_{\Delta}}(r)$, is directly comparable to the usual real-space correlation function,  $\xi (r)$.

To calculate statistical errors, we use jackknife resampling of 39 spatially coherent regions, defined such that each one has  $\sim\!1$,$500$ primary spectroscopic PRIMUS galaxies. Each jackknife region is of order $\sim\!0.25$deg$^{2}$, and is divided in RA and Dec, such that all PRIMUS subfields have between three and sixteen subregions, as shown in Figure 1. The spatial covariance matrix is computed using the same conventions as \citet{Zehavi11}. For overdensities, $\xi_{_{\Delta}}$, at scales $r_{i}$ and $r_{j}$, the covariance between the two values is given as follows:

\begin{equation}
{\rm Cov}( \xi^{}{_{\Delta}}_{i}, \xi^{}{_{\Delta}}_{j}) = \frac{N - 1}{N}\sum\limits_{l = 1}^{N} ( \xi^{}{_{\Delta}}_{i}^{l} - \overline{ \xi^{}{_{\Delta}}}_{i}) ( \xi^{}{_{\Delta}}_{j}^{l} - \overline{ \xi^{}{_{\Delta}}}_{j}),
\end{equation}
where $N = 39$ in this work, and $\overline{ \xi^{}_{_{\Delta}}}_{i}$ is the mean value of all subregions at scale $r_{i}$.

\begin{table*}[t]
\tablewidth{0pt}
\begin{center}
Table 1 \\
PRIMUS Spectroscopic and Photometric Statistics \\
\footnotesize
\begin{tabular}{lcrccccccccc}
\\
\tableline
Field & RA$^{\,a}$ & Dec$^{\,a}$ & A$_{spec}^{\,b}$ & $N_{spec}^{\,c}$ & $N_{red}^{\,d}$ & $N_{green}^{\,d}$ & $N_{blue}^{\,d}$ & $N_{jack}^{\,e}$ & A$_{img}^{\,f}$ &$N_{img}^{\,g}$  \cr
\tableline
\tableline
\hline
CDFS & 03:32 & -28:54 & 1.95 & 12481 & 2726 & 1258 & 8461 & 8 & 2.14 & 74000 &\cr
COSMOS & 10:00 & +02:21 & 1.03 & 7918 & 1802 & 737 & 5379 & 5 & 1.59 & 59000 &\cr
DEEP2 02$hr$ & 02:30 & +00:36 & 0.58 & 4849 & 1030 & 500 & 3319 & 3 & 0.86 & 25000 &\cr
DEEP2 23$hr$ & 23:30 & +00:09 & 0.67 & 4814 & 1261 & 374 & 3179 & 3 & 0.96 & 25000 &\cr
ELAIS S1 & 00:36 & -43:30 & 0.90 & 6676 & 1909 & 823 & 3944 & 4 & 1.41 & 39000 &\cr
XMM-LSS & 02:20 & -04:45 & 2.88 & 25368 & 5740 & 2254 & 17374 & 16 & 15.72 & 322000 &\cr
\tableline
Total   &   &   & 9.05 & 62106 & 14504 & 5946 & 41656 & 39 & 22.68 & 545000 \\
\end{tabular}
\end{center}
\scriptsize
$^{a}\,$ Approximate field centers (J2000); see \citealt{Coil11}) for details \\
$^{b}\,$ Area of primary science fields (in deg$^{2}$); does not include primary objects removed due to location outside of imaging mask \\
$^{c}\,$ Number of primary PRIMUS objects used in this work \\
$^{d}\,$ Number of primary PRIMUS objects in each color sample \\
$^{e}\,$ Number of jackknife regions used to calculate cosmic variance errors\\
$^{f}\,$ Exact area of imaging mask (in deg$^{2}$) \\
$^{g}\,$ Approximate number of imaging galaxies \textit{used} in cross-correlation calculation \\
\end{table*}

\section{Data}
\subsection{PRIMUS Motivation}

The PRIsm MUlti-object Survey (PRIMUS; \citealt{Coil11}; \citealt{Cool13}) is a wide-area, spectroscopic, faint galaxy survey out to $z \sim 1$. Using a low-dispersion prism instrument, PRIMUS obtained robust redshifts of $\sim\!130$,$000$ unique objects to an accuracy of $\sigma_{z}/(1+z) \sim 0.005$ over 9.1 deg$^{2}$ of sky and to a depth of $i_{AB} \sim 23.5$.

To meet these needs, PRIMUS designed a low-dispersion prism to be installed on the Magellan I (Baade) 6.5m telescope at Las Campanas Observatory. It measured redshifts with a resolution of $R = \lambda/\Delta \lambda \sim 40$. This is substantially higher resolution than most photometric redshifts ($\lambda/\Delta \lambda \sim\! 3 - 5$) and still more than a factor of two better than the best photometric surveys, such as COMBO-17 \citep{Wolf04}) or ALHAMBRA \citep{Moles08} . This reduces the average redshift errors, $\sigma_{z}/(1+z)$, from $\sim\! 3 - 5 \%$ to only $0.5 \%$ compared to the usual photometric errors. Moreover, the redshift completeness in PRIMUS is not color-dependent \citep{Cool13}. These errors are sufficient to measure large-scale clustering, and by reducing uncertainty in the distance modulus, they also better constrain the galaxy luminosities --- important if we wish to measure luminosity-dependent clustering.

The IMACS camera on Baade has a effective field of view of 0.18 deg$^{2}$, making it a good choice for a wide-field survey. Compared to a photometric redshift survey, the decision to use a prism meant that otherwise blank sky pixels would instead gather spectral information. The prism also allowed for $\sim\!2$,$500$ objects to be observed simultaneously in each pointing, more than a traditional grism or grating. This multiplexing on a large telescope made possible both the excellent survey depth and the high number of measured redshifts. For a more recent implementation of similar methods, see \citet{Kelson14}.

\subsection{Science Fields and Photometric Catalogs}
The PRIMUS target selection and data reduction pipeline are described in detail by Coil \etal (2011) and Cool \etal (2013). This works uses six of the science fields with PRIMUS redshifts. These are the Chandra Deep Field South-SWIRE field (CDFS, \citealt{Giacconi01}), the European Large Area ISO Survey-South 1 field (ELAIS-S1, \citealt{Oliver00}), the DEEP2 02$^{hr}$ and DEEP2 23$^{hr}$ fields, the COSMOS field (\citealt{Scoville07}), and the XMM-Large Scale Structure Survey field (XMM-LSS, \citealt{Pierre04}).

For CDFS, we used SWIRE photometry \citep{Lonsdale03}.  For ELAIS-S1, we used combined photometry from ESO/WFI and VLT/VIMOS (\citealt{Berta06}, 2008). For XMM-LSS, we used the photometry from the $CFHT$ Legacy Survey \citep{Coupon09} that has been reprocessed and published by \citet{Erben09}. For the DEEP2 fields, we use their photometry from fields 3 and 4, which are the 23$^{hr}$ fields and  02$^{hr}$ fields, respectively \citep{Coil04}. For COSMOS, we used the April 2009 data release \citep{Ilbert09}. Figure 1 shows the sky coverage of the spectroscopic and imaging samples for each science field. Table 1 lists properties of each field for both the spectroscopic and imaging samples, whose selections are described in detail in the subsequent section.

\subsection{Imaging Sample Selection}
We select our secondary imaging sample from the parent catalogs in two steps. First, as necessary, we apply zero-point corrections, convert from Vega to AB magnitudes, apply extinction corrections, and remove stars using star-galaxy flags as given by each survey. See \citet{Coil11} for further details on these corrections, which are identical in our science analysis, except for CDFS zeropoints, for which no offset was applied. We remove any objects that fall outside of our imaging masks. It is imperative that the photometry inside our masks be as uniform as possible (e.g., surveyed to the same depth) and that it not be affected by cosmic rays, bad pixels, or saturation due to nearby bright stars. Thus, we choose to be relatively conservative in expanding bright star regions and masking areas in which there is uneven coverage (e.g., vignetting at survey edges). Our survey region boundaries are defined using \textbf{\texttt{mangle}} (v2.2, \citealt{Swanson08b}). 

Second, for the entire imaging sample, we precompute a grid of k-corrections from the entire range $0.2 \leq z \leq 1.0$ in increments of $\Delta z = 0.02$. This reduces the computational overhead of selecting the secondary luminosity-defined tracer galaxies. We assign to each imaging galaxy the absolute magnitude it would have at the redshift of the spectroscopic galaxy. Then, we apply the appropriate k-correction (\textbf{\texttt{kcorrect}}, version 4.2; \citealt{BR07}) and passively evolve the galaxy from this redshift to $z = 0.1$, by linearly interpolating between the grid of k-corrections and evolving the imaging galaxy as $M_{g,0.1,k+e} =  M_{g,0.1,k} + Q(z - 0.1)$, with $Q = 2.04$ magnitudes per unit redshift \citep{Blanton03}. A photometric galaxy becomes part of the secondary sample for a given primary object if $M_{g,0.1,k+e}$ falls within an empirically defined range. An empirical tracer sample is defined as:

\begin{equation}
M_{g,0.1,k+e}^{*} - 0.5 \le M_{g,0.1,k+e} \le M_{g,0.1,k+e}^{*} + 1.0
\end{equation}
where $M_{g}^{*} = -19.39$ \citep{Blanton03}. Note that, unless otherwise indicated, all absolute magnitudes in the paper are k-corrected and passively evolved to z=0.1, but henceforth, we will suppress the additional subscripts, and write the absolute magnitudes more concisely as $M_{g}$.

Choosing a fixed absolute magnitude range for our imaging sample defines a nearly uniform population of tracers galaxies. The actual spatial density of the imaging catalog at a particular redshift is unknown because it depends on the true evolution of a diverse population of galaxies that includes a range of spectral and morphological types. If our model evolution differs from the actual evolution, then our clustering amplitude measurement will evolve anomolously in redshift, even if the underlying clustering amplitude of our primary spectroscopic sample is constant; hence, we do not attempt to constrain redshift evolution in this work. Nonetheless, the relative clustering amplitudes of different luminosity and color subsamples at a given redshift are still exactly correct, because the same secondary tracer population is used to measure the cross-correlation around all PRIMUS galaxies, and these relative amplitudes (e.g., red vs. blue clustering) can be compared exactly across redshifts.

While we have attempted to reduce the number density evolution of our secondary sample, we are constrained by completeness limits in the imaging catalogs at high redshift. A lower value of passive evolution, Q, for our empirical selection would bring the SDSS results into better agreement with, for example, the DEEP2 luminosity function \citep{Willmer06} at intermediate redshift. However, using the higher passive evolution measured by SDSS allows us to use the faintest possible selection at high redshift, while still using $\sim L^*$ tracers at low redshift. This implies that we are likely cross-correlating our PRIMUS galaxies with a slightly more biased tracer population at higher redshifts than at lower redshifts. We account for this offset in our calculation of the real-space number densities of our tracer sample. We integrate the luminosity function to obtain the number density for our empirically-selected tracer population. We choose the theoretical luminosity evolution model so that the predicted number density at $z = 0.5$ of a 1.5 magnitude bin around $M^{*}$ is roughly equivalent given the luminosity functions of SDSS \citep{Blanton03} and DEEP2 \citep{Willmer06}. We used the DEEP2 fits for their $\langle z \rangle = 0.9$ sample with `optimal' weights. This is still necessarily an approximation because the luminosity functions do not agree on the slope of the faint end. Table~2 gives the values of $\phi_{com}$ in increments of $\Delta z=0.05$. 

The theoretical evolution model passive evolution model is:

\begin{eqnarray}
M_{g,0.1,k+e} & = & M_{g,0.1,k} + Q(z - 0.1)  \\
Q & = & q_{0}(1 + q_{1}(z - 0.1))
\end{eqnarray}
where $q_{0} = 2.0$ and $q_{1} = -0.8$.

\subsection{Spectroscopic Sample Selection}
We select our spectroscopic sample from PRIMUS's `primary' catalog, whose known selection function allows us to create a statistically complete galaxy sample. We select only those galaxies with high-quality redshifts ($Q \ge 3$) in the range $0.2 \leq z \leq 1.0$.  We have tested our analysis on only the highest redshift quality, $Q4$, selection, and the difference in results is within our statistical errors, so we choose to use the somewhat larger sample. From comparisons to higher-resolution spectroscopy, the fractional redshift error $\Delta_{z}/(1+z) < 0.03$ for $92.2\%$ galaxies and  $< 0.1$ for $98\%$ of galaxies. The dispersion among the $92.2\%$ with the highest quality redshifts is $\sigma_{z}/(1+z) \sim 0.0051$ (for details, see \citealt{Coil11}). As noted in Section 2, each galaxy in the \textit{primary} catalog has an individual completeness weight by which we weight all of our clustering measurements. We eliminate from our primary sample any galaxy that falls outside our imaging mask. This affects at most only a handful of objects in all fields except CDFS, where it affects $\sim\! 600$ galaxies. Even then, this is only $\sim 1\%$ of our primary catalog, and so we err on the side of being conservative and avoid galaxies where photometric issues might bias our results. Lastly, we exclude any primary galaxies for which the angular masks account for more than $50\%$ of the projected area on the sky at the largest scale.

In this paper, we report results in terms of the k-corrected and passively evolved absolute magnitudes, $M_{g}$, of the PRIMUS galaxies. We use the quadratic passive evolution model given in the previous subsection, where $q_{0} = 2.0$ and $q_{1} = -0.8$. Figure 2 plots absolute magnitude versus redshift for our primary sample.

\begin{figure}[t]
\plotone{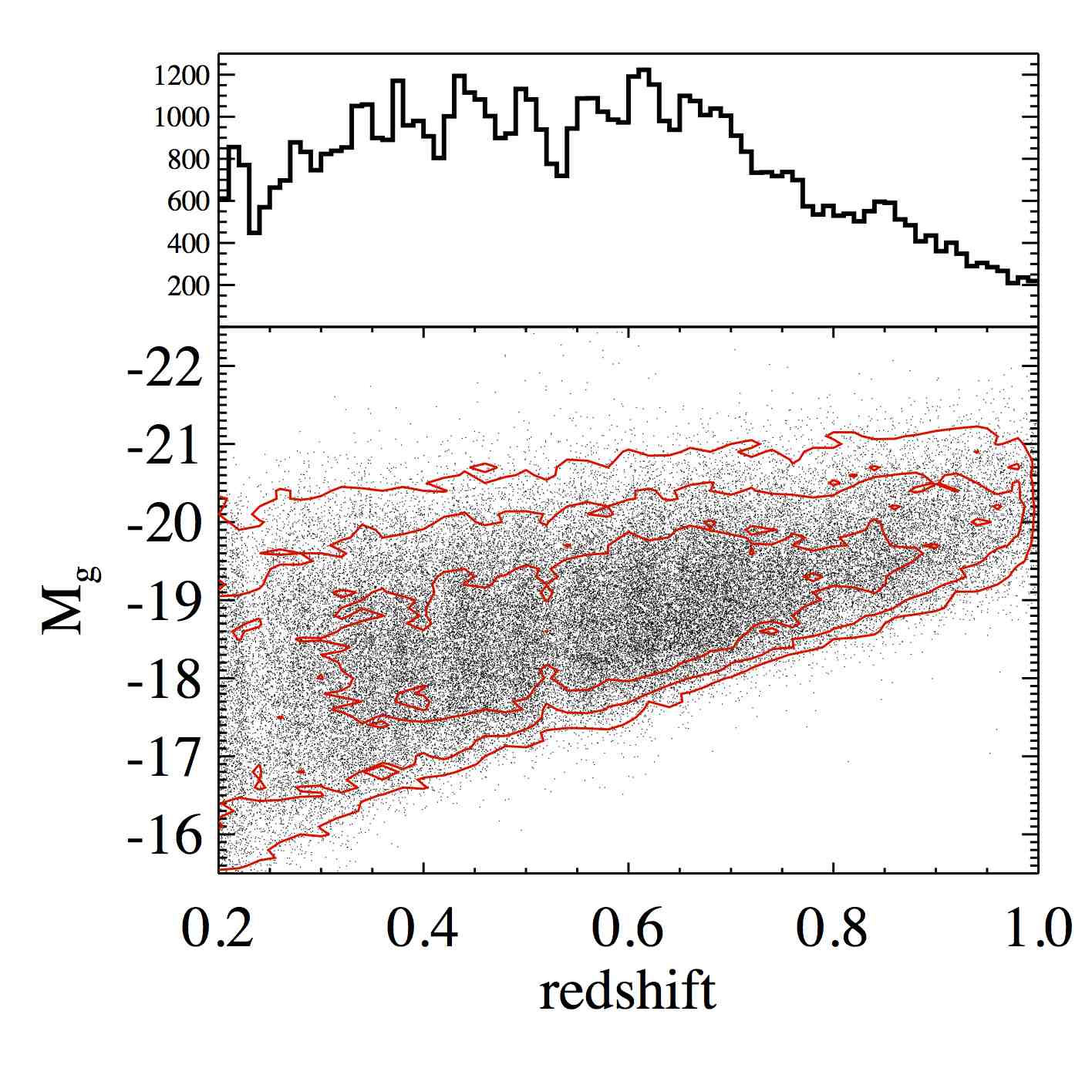}
\caption{\label{fig:fig2}
\small 
Contours for $50\%$, $87\%$, and $99.7\%$ inclusion in absolute $M_{g}$ versus redshift space for the $62$,$106$ PRIMUS galaxies used in this work. Galaxies are k-corrected and passively evolved to $z = 0.1$. Top: A histogram of the redshifts in $\Delta z = 0.1$ bins.
}
\end{figure}

\begin{figure}[t]
\plotone{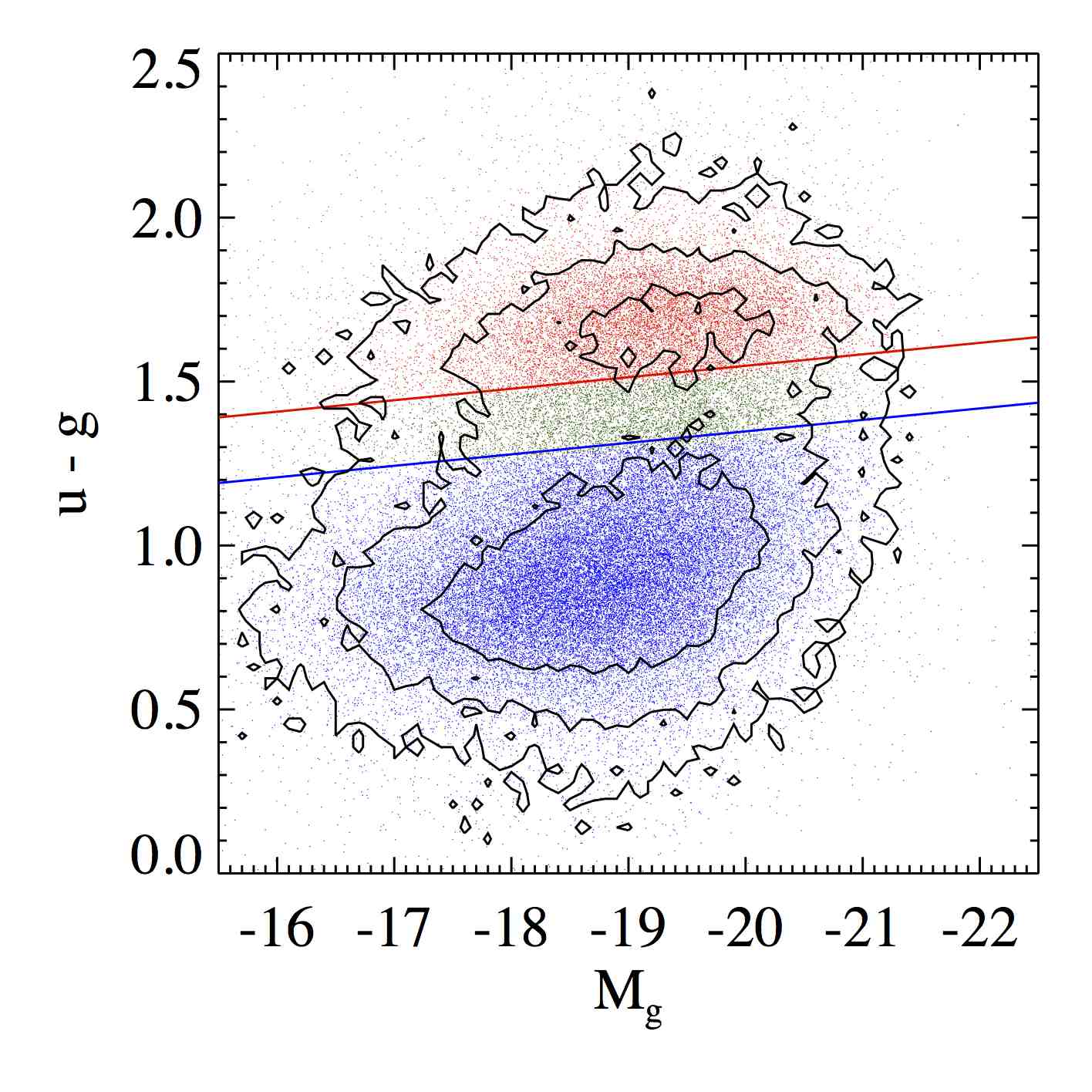}
\caption{\label{fig:fig3}
\small 
Inclusion contours, as in Fig. 2, but in the color-magnitude diagram of primary galaxies. Solid lines show color cuts between the red, green, and blue samples, as defined by Eqns. $(12)$ and $(13)$, for a hypothetical galaxy at $z = 0.5$, but actual color bins are redshift dependent. See Table 1 and Section 3.4 for selection details.}
\end{figure}

\begin{table}[t]
\tablewidth{0pt}
\begin{center}
Table 2 \\
Modeled Densities of Secondary Sample  \\
\footnotesize
\begin{tabular}{ccccccc}
\cr
\tableline
$z$ & $\phi_{com} (z)^{\,a}$ & $z$ & $\phi_{com} (z)^{\,a}$ & $z$ & $\phi_{com} (z)^{\,a}$ &  \cr
\tableline
\tableline
0.20 & 1.2798 & 0.50 & 1.0741 & 0.80 & 0.6306 &\cr
0.25 & 1.2623 & 0.55 & 1.0155 & 0.85 & 0.5424 &\cr
0.30 & 1.2384 & 0.60 & 0.9499 & 0.90 & 0.4545 &\cr
0.35 & 1.2077 & 0.65 & 0.8779 & 0.95 & 0.3693 &\cr
0.40 & 1.1702 & 0.70 & 0.7999 & 1.00 & 0.2649 &\cr
0.45 & 1.1257 & 0.75 & 0.7171 &\cr 

\end{tabular}
\end{center}
\scriptsize
$^{a}\,$ The values of the comoving densities $\phi_{com} (z)$, which are calculated by matching SDSS \citep{Blanton03} and DEEP2 \citep{Willmer06} luminosity functions at z=0.5, are in units of $10^{-2} \hmpcC$. Because our measurements are in physical units, when we calculate $\xi_{_\Delta}$, we divide out the proper density, $\phi_0 =  (1+z)^3 \phi_{com}$.\\
\end{table}

We divide our primary science sample into three rest-frame color bins: red, green, and blue. We do so using a linear fit to the red sequence, using $M_{g}$ and $u - g$ color, over the full redshift range of our primary sample, and over the magnitude range $-21 \leq M_{g,k} \leq -17$, where the red sequence is well-sampled. We fit using k-corrected (but not passively evolved) absolute magnitudes. Specifically, the $M_{g}$ vs. $u - g$ color space is binned in 0.25 increments in magnitude and 0.05 in color, and the red sequence is defined in each magnitude bin as the peak of the color distribution. We fit the red sequence slope, and then shift it in color-space to define the other color bins. The shift is chosen so that the color cuts roughly match those of \citet{Skibba14}, in which magnitudes were k-corrected to $z = 0$ rather than to $z = 0.1$, so that the results can be more easily compared. We then allow for redshift evolution in the color cuts using the values found by \citet{Aird12}. We obtain the following color cuts for the lower bound of the red sample and the upper bound of the blue sample:

\begin{eqnarray}
(u - g)_{red} & = & - 0.035M_{g} - 0.065z + 0.848 \\
(u - g)_{blue} & = & - 0.035M_{g} - 0.065z + 0.648
\end{eqnarray}

The green sample is the independent subset from the upper blue boundary to the lower red boundary. Figure 3 plots $M_{g}$ versus $u - g$ color for our spectroscopic sample, with the color cuts for a galaxy at $z = 0.5$ overlaid.

Having chosen to use a flux-limited sample to enhance the size of our primary sample, we need to account for the redshift-dependence in out results. We do this in two ways, each of which highlights different science results. First, we fit the clustering amplitudes to a joint power-law model in redshift and luminosity. For each color sample, the model takes the form:

\begin{equation}
\xi^{data}_{_{\Delta, i}}= \xi^{model}_{_{\Delta}} \left(\frac{1 + z}{1.50}\right)^{\alpha} \left(\frac{L}{1.5 L^{*}}\right)^{\beta_{i}}
\end{equation}
where $\xi^{model}_{_{\Delta}}$ is the best-fit amplitude at the fiducial normalization of $z = 0.50$ and $L = 1.5 L^{*}$ at the \textit{i}th scale. We fit the model individually at each proper scale over a dense grid in redshift while allowing only the amplitude and $\beta_{i}$ to vary, and then we calculate the best-fit using $\chi^2_{total} = \sum\limits_{i = 0}^{i = 4} \chi^2_i$, for each grid point in redshift. By this, we demand that the redshift evolution of the clustering amplitude be the same as all physical scales, but we allow the luminosity dependence to vary as a function of scale. The normalizations at $z = 0.50$ and $L = 1.5 L^{*}$ are chosen at convenient locations near the median redshift value of the overall primary sample, such that small variations in these normalizations have negligible effect on the best-fit clustering amplitude $\xi^{model}_{_{\Delta}}$. We suppress the superscript in our plots.

Second, we present empirical clustering measurements without any underlying assumption that the luminosity dependence is a power-law. However, because we are using a flux-limited sample, the mean redshifts of brighter magnitude bins are higher than those of lower magnitude bins by approximately $z \sim 0.1$ per magnitude. Lest we confuse redshift evolution for luminosity dependence, we ``detrend" the empirical clustering amplitudes on a object-by-object basis before binning them in magnitude. Specifically, we divide the measured $\xi^{data}_{_{\Delta}}$ by $\left( \frac{1 + z}{1.50}\right)^{\alpha}$, where we use the color-dependent value of $\alpha$ calculated using the above model. Thus, the clustering amplitudes presented in this way once again represent the clustering strength of a fiducial galaxy at $z = 0.5$.

\begin{figure}[t]
\plotone{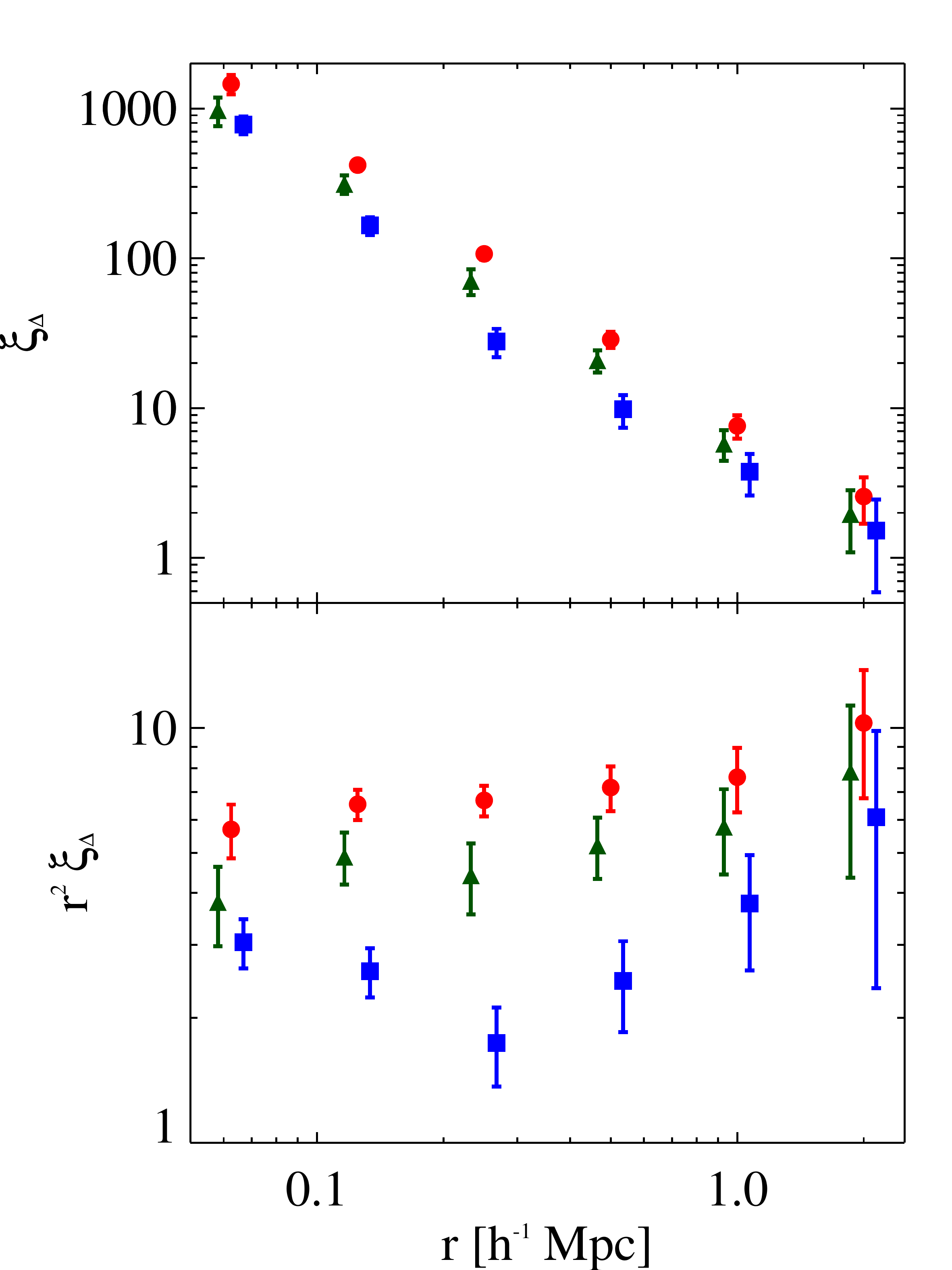}
\caption{\label{fig:fig4}
\small 
Modeled overdensity ($\xi_{_{\Delta}}$) around red, green, and blue galaxies (red circles, green triangles, and blue squares, respectively) as a function of scale $\hmpc$. The secondary sample has a redshift-evolving 1.5 magnitude range, as given by Eqn.~$(9)$. The model is a joint power-law fit to redshift and luminosity dependence, where redshift dependence is fixed across scales, but luminosity dependence varies independently at each scale. The clustering amplitude is normalized at $z = 0.5$ and $L = 1.5 L^{*}$. The top panel plots $\xi_{_{\Delta}}$, whereas the bottom panel plots $r^{2}\xi_{_{\Delta}}$, in order to better resolve the fluctuations with scale. Values on the horizontal axis are offset by $7\%$ for visual clarity. All error bars represent jackknife resampling over 39 spatially contiguous regions, as shown in Fig. 1.
}
\end{figure}

\begin{figure}[t]
\plotone{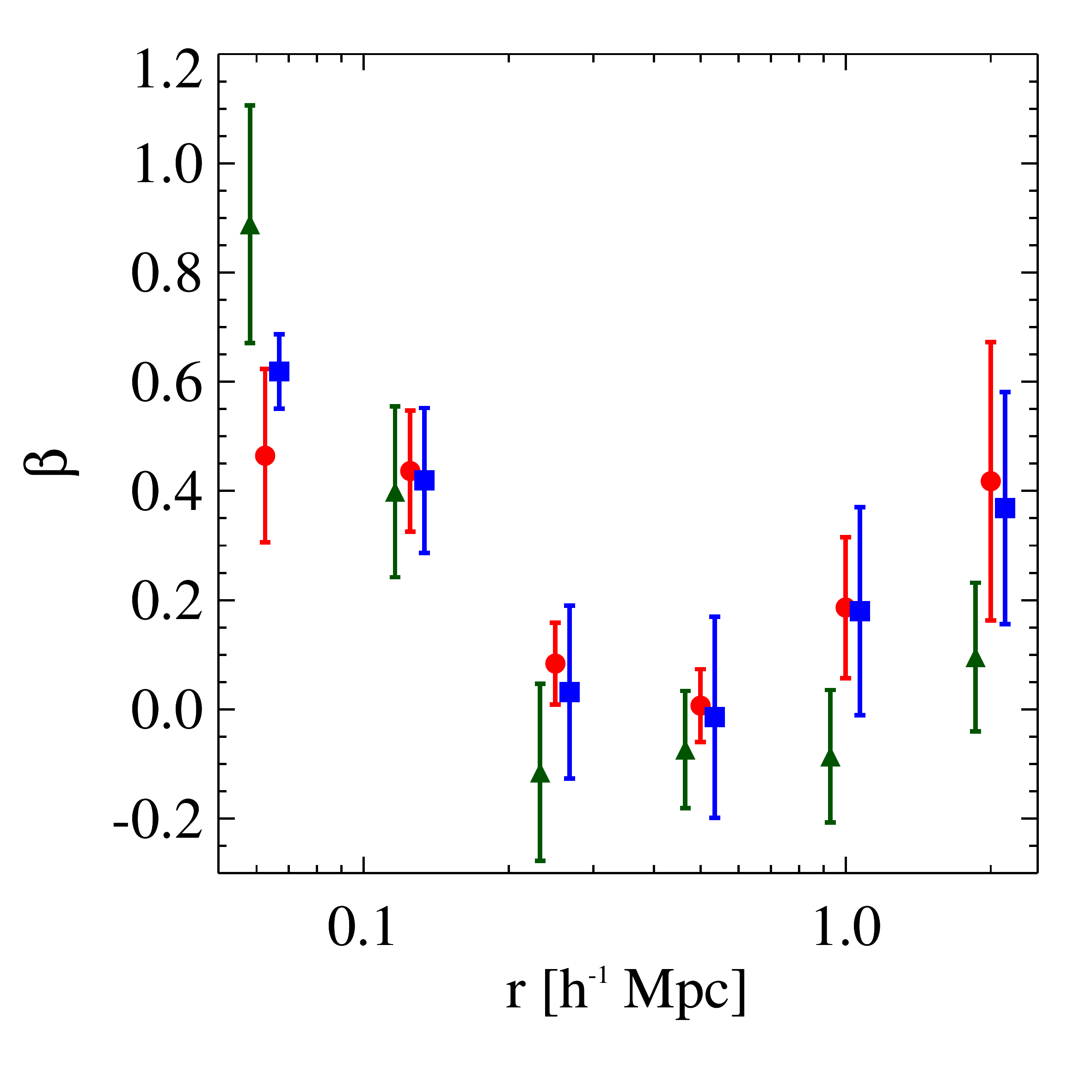}
\caption{\label{fig:fig5}
\small 
The best power-law exponent fit for luminosity ($\beta$) as a function of scale for the amplitudes shown in Fig.~4.   Red, green, and blue galaxies are represented by red circles, green triangles, and blue squares, respectively. The functional form of the fit is given by Eqn.~$(14)$, and the jointly best-fit redshift evolution (which is not scale dependent) is $\alpha_{red} = -1.90 \pm 0.64$, $\alpha_{green} = -1.25 \pm 0.92$, and $\alpha_{blue} = -2.10 \pm 1.15$. Luminosity dependence is strongest at smaller scales for all three color samples.
}
\end{figure}

\section{Results}

In Figure~4 (top), we present our best-fit cross-correlation measurements $\xi_{_{\Delta}}$ of the primary PRIMUS galaxies with respect to the photometric galaxies in the fixed, k+evolution corrected, absolute magnitude range $M^*_g - 0.5 < M < M^*_g + 1$. These clustering amplitudes correspond to the fiducial value $L = 1.5L^*$ and $z = 0.5$, as given in Equation 13. The results cover proper scales from $r = 0.0625 \hmpc$ to $r = 2.0 \hmpc$, and we split them according to the color samples defined in Section 3. The bottom plot presents $r^2 \xi_{_{\Delta}}$ so as better to display small deviations from a power-law. Note that $M^*_g \approx -19.93$ at $z = 0.5$ using our quadratic passive evolution model. The corresponding best-fit values for the redshift evolution is $\alpha_{red} = -1.90 \pm 0.64$, $\alpha_{green} = -1.25 \pm 0.92$, and $\alpha_{blue} = -2.10 \pm 1.15$, and the corresponding best-fit luminosity dependence are shown in Figure 5. We highlight key points in this section, and we discuss them in light of theory and other observations in the following section.

The real-space cross-correlation function of PRIMUS galaxies shows a strong dependence on color. The clustering amplitude of red galaxies is $\sim\!1.5$ to $3$ times that of blue galaxies and $\sim 1$ to $1.5$ times that of green galaxies. Given the statistical errors, this relative clustering bias between red and blue galaxies and between red and green galaxies appears mostly constant with physical scale. However, at $250 \hkpc$, where the blue galaxy sample displays a noticeable inflection from a power-law, there is an increase in the relative bias of red and blue galaxies. Table 4 reports the ratios of the red-green and red-blue clustering amplitudes as a function of scale. 

In Figure~5, we show the best-fit luminosity power-law exponents as a function of proper scale.  We find clear evidence that galaxies exhibit increasingly strong luminosity dependence at the smallest scales, regardless of color, whereas luminosity dependent clustering is consistent with zero at $250 \hkpc$ and $500 \hkpc$. The red and blue galaxy samples again appear to show slight luminosity dependence at $1 \hmpc$ and above, but the statistical errors are large.  However, as shown next, a power-law model of the luminosity dependence is not sufficient to explain the variation in clustering as a function of luminosity. Values for the best-fit $\xi_{_{\Delta}}$ and luminosity dependence parameter ($\beta$) --- along with their respective jackknife errors --- are given in Table~3.

Figure~6 presents $r^2 \xi_{_{\Delta}}$ as a function of magnitude. Luminosity dependence appears more complicated than a simple power-law relation, and so unlike in the Figure~4, we show the empirical averages of the clustering amplitudes, not fits. However, because we are using a flux-limited sample, the mean redshifts of the brighter magnitude bins are higher than the lower magnitude bins by approximately $z \sim 0.1$ per magnitude, and so as noted in Section~3, we detrend the galaxies to our fiducial redshift of $z= 0.5$. Due to Poisson fluctuations and the large number of bins created by subdividing in both color and luminosity, there are several bins in which the clustering is measured to be negative. We plot those points at $10^{-4}$ instead, so that the $1 \sigma$ errors can be viewed on the logarithmic scale.

The results indicate that the luminosity dependence is not smooth over the range of luminosities available in PRIMUS. While blue galaxies' luminosity dependence appears to be fairly well-approximated by a power-law, the red galaxies have had their luminosity dependence underestimated by fitting a power-law. In particular, the apparent lack of luminosity dependence at $250 \hkpc$ and $500 \hkpc$ in Figure~5 obscures non-monotonicity in the luminosity dependence. On these scales, red galaxies show the weakest clustering at roughly $L^{*}$, with strongly increasing clustering at higher luminosities, and weaker increases in clustering strength at fainter luminosities. The green galaxy sample, while noisy, displays similar clustering strength to the blue sample at the smallest scales, but is indistinguishable from the red galaxy sample at larger scales.

\begin{table*}[t]
\tablewidth{0pt}
\begin{center}
Table 3 \\
Clustering and Luminosity Fits by Scale and Color for Fiducial $z = 0.5$ and $L = 1.5 L^*$ \\
\footnotesize
\begin{tabular}{lrrrrrrr}
\cr
\tableline
& $0.0625 \hmpc$ & $0.125 \hmpc$ & $0.250 \hmpc$ & $0.50 \hmpc$ & $1.00 \hmpc$ & $2.00 \hmpc$ & \cr
\tableline
\tableline
& & & $\xi_{_{\Delta}}^{model}$& & & \\[3pt]
\tableline

Red & $1457 \pm 215$ & $419 \pm 35$ & $107 \pm 9$ & $28.7 \pm 3.6$ & $7.61 \pm 1.35$ & $2.57 \pm 0.88$ &\cr
Green & $973 \pm 211$ & $313 \pm 45$ & $70.6 \pm 13.7$ & $20.8 \pm 3.5$ & $5.78 \pm 1.34$ & $1.96 \pm 0.87$ & \cr
Blue & $780 \pm 106$ & $166 \pm 23$ & $27.8\pm 6.0$ & $9.82 \pm 2.43$ & $3.77 \pm 1.17$ & $1.52 \pm 0.93$ & \cr
\tableline
& & & $\beta$& & & \\[3pt]
\tableline

Red & $0.46\pm 0.16$ & $0.44 \pm 0.11$ & $0.084 \pm 0.075$ & $0.007 \pm 0.066$ & $0.19 \pm 0.13$ & $0.42 \pm 0.25$ &\cr
Green & $0.89 \pm 0.22$ & $0.40 \pm 0.16$ & $-0.12 \pm 0.16$ & $-0.07 \pm 0.11$ & $-0.086 \pm 0.12$ & $0.096 \pm 0.14$ &\cr
Blue & $0.62 \pm 0.07$ & $0.42 \pm 0.13$ & $0.03 \pm 0.16$ & $-0.01 \pm 0.18$ & $0.18 \pm 0.19$ & $0.37 \pm 0.21$ &\cr
\tableline

\end{tabular}
\end{center}
\end{table*}

\begin{figure*}[t]
\plotone{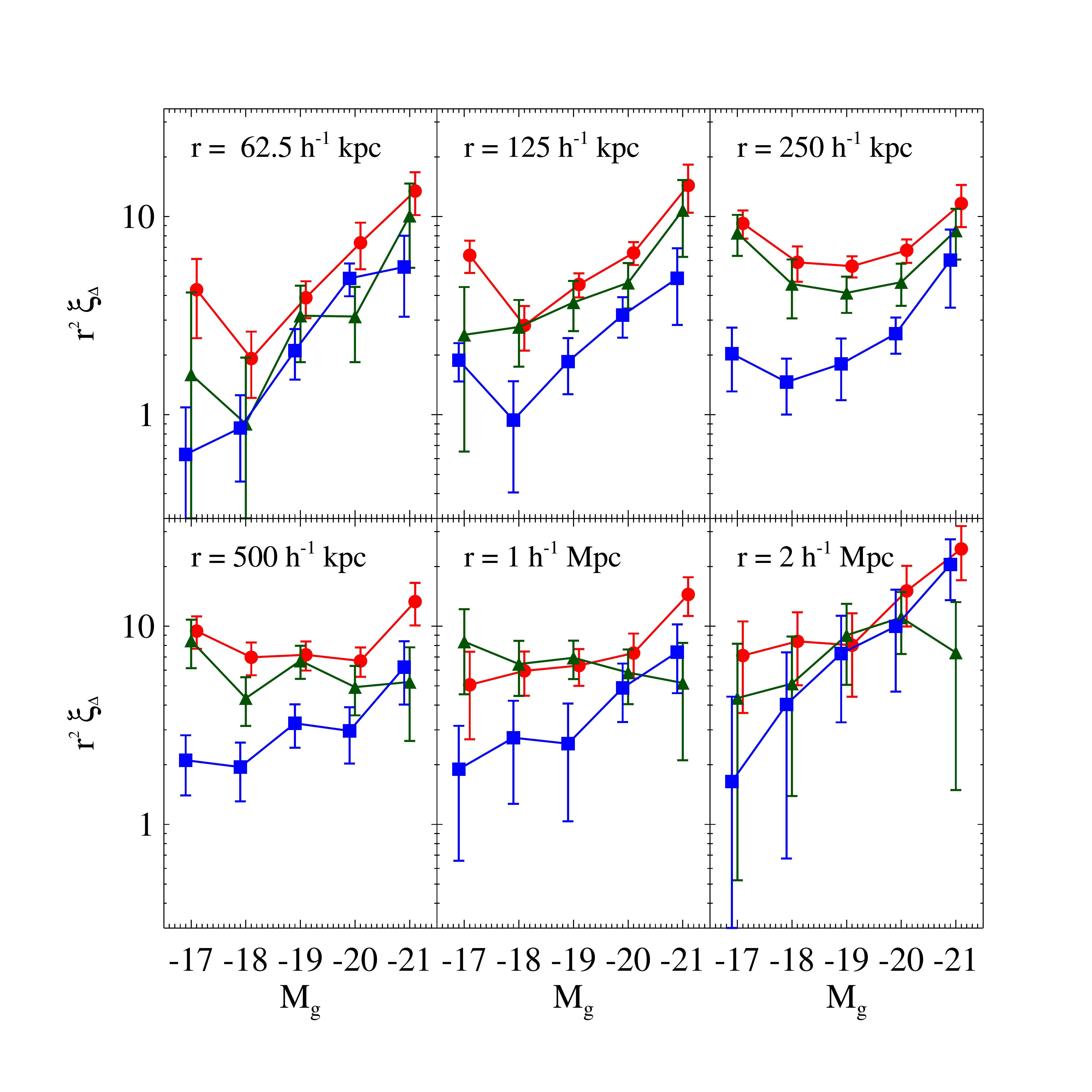}
\caption{\label{fig:fig6}
\small 
Overdensities ($\xi_{_{\Delta}}$) for the red, green, and blue sample (red circles, green triangles, and blue squares, respectively) as a function of luminosity $M_g$. Bins are each one magnitude in width, and centered on integer magnitudes from $M_g = -17$ to $M_g = -21$. Significant breaks from a single power-law in luminosity can be seen, particularly for red galaxies. Measurements have been normalized to redshift $z = 0.5$; see Sec. 3.4 for details. 
}
\end{figure*}

\section{Discussion}

\subsection{Comparison to Past Studies}

Our results present a methodologically unique look at galaxy clustering at a fiducial redshift of $z = 0.5$. In particular, we use cross-correlations to extend the study of luminosity and color dependence down to $62.5 \hkpc$, which is smaller than other works in the literature that look at similar trends (e.g., \citealt{Hogg03}, \citealt{Coil08}, \citealt{Zehavi11}, \citealt{Marulli13}, \citealt{Skibba14}). Because our results examine scales that are primarily a measure of the one-halo clustering, measurement of the clustering length or clustering slope are not readily comparable with those measured using projected statistics to much larger scales. However, we can still qualitatively compare the clustering as a function of color, luminosity, and scale.

In our best-fit amplitude measurements of color-dependent clustering in Fig.~\ref{fig:fig4}, we find an offset of $\sim 2~\rm{to}~3$ between red and blue samples at $L = 1.5 L^*$ in the cross-correlation amplitude, which would be equivalent to an offset of $\sim 4~\rm{to}~9$ in the auto-correlation function. This is consistent with both low redshift SDSS results (\citealt{Zehavi11}) and high redshift DEEP2 results (\citealt{Coil08}). We find that this relative clustering offset of red and blue galaxies continues to $62.5 \hkpc$ without any noticeable increase in magnitude. In fact, the largest relative clustering difference as a function of scale occurs at $r = 250 \hkpc$, which we attribute to the fact the transition between one- and two-halo clustering terms occurring at smaller scales for blue than for red galaxies, due to their location in less massive halos at fixed luminosity. As seen in Fig.~\ref{fig:fig6}, this feature appears driven primarily by the fainter blue galaxies, but is still present in all but the highest luminosity bin. \citet{Skibba14} also see a small but significant increase in the red to blue bias for their lowest luminosity threshold sample at the same scale (see their Fig.~11).

We also find a monotonic increase in clustering strength from blue to red galaxies, such that green galaxies have an intermediate clustering strength. In both Fig.~\ref{fig:fig4} and Fig.~\ref{fig:fig6}, we see some evidence that the green galaxy population may be a transitional population, undergoing quenching via the environment or internally, rather than being simply a mixture of overlapping red and blue galaxy distributions (e.g., \citealt{Mendez11}). At all scales, the clustering of green galaxies is statistically indistinguishable, but almost exclusively lower, than that from the red galaxies. However, at scales of $r = 250 \hkpc$ to $r = 1 \hmpc$, the blue population demonstrates a significantly lower clustering signal than either the red of the green samples, whereas at the smallest scales, there is no statistical difference between the green and blue clustering signal. Thus, it is possible that at large scales, green galaxies are demonstrating clustering similar to the red galaxies, while at smaller scales showing clustering similar to the blue galaxies. A similar result was previously found by \citet{Coil08} at $z \sim 1$, but not by \citet{Zehavi11} at $z \sim 0$. Alternately, the green population may be substantially identical to the red population, but mixed with some blue galaxies. Most directly, however, the color-dependent results in the projected auto-correlation measured by \citet{Skibba14} in PRIMUS do not appear to show this scale dependence for the green valley (see their Fig.~14). This apparent discrepancy could be a result of the combination of different color bins and different scales. \citet{Skibba14} measure substantial difference between the `redder' and `reddest' samples, but little difference between the `redder' and `green' samples. Furthermore, the projected statistics mix real-space scales, so the the convergence of blue and green clustering results seen in our work occurs at scales not adequately probed by \citet{Skibba14}. Thus we cannot reliably discern between the two possibilities, not only because we are limited by small number statistics (see Table~5) but particularly because at the smallest scale the overall difference in clustering amplitude between red and blue decreases relative to the difference at $r = 250 \hkpc$. Nonetheless, this points to the impact such selection effects can have and motivates studying clustering as a function of intrinsic physical, rather than observable, properties, such as that being done by \citet{MendezPREP}, which is examining PRIMUS clustering as a function of specific star formation rate and stellar mass.

As shown in Fig.~\ref{fig:fig5}, we find roughly equal luminosity dependence in all of our color samples. For $r \geq 250 \hkpc$, Fig.~\ref{fig:fig6} shows luminosity dependence in the red population only our brightest bin. The lack of luminosity dependence at scales of roughly $r = 250 - 500 \hkpc$ confirms other observations at higher redshift. VIPERS \citep{Marulli13} also shows a lack of either luminosity or stellar mass dependence at this scale in their redshift-space correlation function (see their Figure 3), while finding more dependence at both smaller and larger physical separations. This effect is washed out in their $w_p$ measurements, so we should not expect to see a similar effect in \citet{Skibba14}, and we do not. On the other hand, \citet{Meneux09} report finding no statistically significant luminosity evolution in the zCOSMOS survey and instead report a trend with stellar mass, particularly at small scales ($r_{p} < 0.3 \hmpc$). Nonetheless, their Fig.~10 shows weak (if insignificant) luminosity evolution at small scales, a trend our results confirm. Moreover, their measurement of stellar mass versus clustering amplitude in their $z = 0.5-0.8$ bin (see their Fig.~15) shows a similar `pinching' at $r_{p} \sim 0.3 \hmpc$, in which there is no trend in stellar mass at that scale, but there is a trend at both lower and higher scales. Moreover, in this redshift bin, \citet{Meneux09} report a one-halo clustering slope that is steeper for the higher stellar mass bins, as well as significant excess power in the two-halo term. They discuss the possibility that these effects are due to known large-scale structure in the COSMOS field at $z \sim 0.73$. If so, then our results, which include the COSMOS field, would be influenced by the same structure. We tested the effect of removing the COSMOS field entirely from our analysis, but while there were minor shifts in clustering amplitude, the luminosity trend at small scales, for all color samples, remained the same.

The luminosity dependence that we find for red galaxies agrees well with \citet{Zehavi11} at low redshift, who find that for large scales, there is little luminosity dependence on the red sequence until $L > 4L^*$, which indeed would enter in our brightest magnitude bin. Likewise, at our smallest scales, we see an increase in clustering among red galaxies. But the increase is not as substantial as in \citet{Zehavi11} --- the clustering remains lower for the faint red galaxies than for our bright red galaxies --- and we only find evidence at the smallest scales, whereas \citet{Zehavi11} observe the increase in clustering out to $r_{p} \approx 2 \hmpc$. On the bright end, again, at intermediate redshifts ($0.20 < z < 0.44$), E05 found a monotonic scale dependence of the luminosity trend across a range of scales from $r \approx 0.22 - 7 \hmpc$ in a sample of LRGs, with smaller scales showing higher luminosity dependence. Meanwhile, at scales of $r \gtrsim  1.75 \hmpc$, the luminosity trend is much less steep for galaxies with $L < 3L^{*}$, while at smaller scales the luminosity dependence was similar regardless of the luminosity range --- results with which PRIMUS concurs at higher redshift.

\begin{table*}[t]
\tablewidth{0pt}
\begin{center}
Table 4 \\
Overdensities Relative to Red Sample from Fits\\
\small
\begin{tabular}{lccccccc}
\cr
\tableline
& $0.0625 \hmpc$ & $0.125 \hmpc$ & $0.250 \hmpc$ & $0.500 \hmpc$ & $1.00 \hmpc$ & $2.00 \hmpc$ &  \cr
\tableline
\tableline
$\xi{_{\Delta,}}_{red}/\xi{_{\Delta,}}_{blue}$ & $1.87 \pm 0.38$ & $2.53 \pm 0.40$ & $3.84 \pm 0.89$ & $2.93 \pm 0.81$ & $2.02 \pm 0.72$ &  $1.69 \pm 1.18$ & \\
$\xi{_{\Delta,}}_{red}/\xi{_{\Delta,}}_{green}$ & $1.49 \pm 0.39$ & $1.34 \pm 0.22$ & $1.52 \pm 0.32$ & $1.38 \pm 0.29$ & $1.32 \pm 0.38$ &  $1.31 \pm 0.73$ & \\[3pt]
\tableline
\end{tabular}
\end{center}
\end{table*}

\begin{table*}[t]
\tablewidth{0pt}
\begin{center}
Table 5 \\
Clustering Results by Magnitude ($r^2 \xi_{_{\Delta}}$)\\
\footnotesize
\begin{tabular}{lrrrrrrrrrr}
\cr
\tableline
$M_g^{max}$ & $M_g^{min}$ & Color & $N_{gal}$ & $0.0625 \hmpc$ & $0.125 \hmpc$ & $0.250 \hmpc$ & $0.50 \hmpc$ & $1.00 \hmpc$ & $2.00 \hmpc$ & \cr
\tableline
\tableline

& &Red &716& $  4.27 \pm  1.84$ & $  6.38 \pm  1.18$ & $  9.24 \pm  1.50$ & $  9.45 \pm  1.76$ & $  5.06 \pm  2.37$ & $  7.11 \pm  3.46$ & \\
-16.5 & -17.5 &Green &372& $  1.60 \pm  2.55$ & $  2.53 \pm  1.88$ & $  8.27 \pm  1.93$ & $  8.46 \pm  2.32$ & $  8.36 \pm  3.82$ & $  4.34 \pm  3.82$ & \\
& &Blue &4357& $  0.63 \pm  0.46$ & $  1.89 \pm  0.41$ & $  2.03 \pm  0.72$ & $  2.11 \pm  0.71$ & $  1.90 \pm  1.24$ & $  1.64 \pm  2.77$ & \\
\tableline

& &Red &2725& $  1.92 \pm  0.71$ & $  2.82 \pm  0.72$ & $  5.88 \pm  1.19$ & $  6.97 \pm  1.31$ & $  5.95 \pm  1.49$ & $  8.39 \pm  3.36$ & \\
-17.5 & -18.5 &Green &1338& $  0.90 \pm  1.04$ & $  2.77 \pm  1.02$ & $  4.57 \pm  1.51$ & $  4.33 \pm  1.20$ & $  6.44 \pm  1.99$ & $  5.13 \pm  3.74$ & \\
& &Blue &12086& $  0.86 \pm  0.40$ & $  0.94 \pm  0.53$ & $  1.46 \pm  0.46$ & $  1.95 \pm  0.64$ & $  2.74 \pm  1.47$ & $  4.03 \pm  3.36$ & \\
\tableline

& &Red &5621& $  3.89 \pm  0.82$ & $  4.54 \pm  0.63$ & $  5.62 \pm  0.69$ & $  7.17 \pm  1.20$ & $  6.33 \pm  1.34$ & $  8.02 \pm  3.61$ & \\
-18.5 & -19.5 &Green &2186& $  3.16 \pm  1.32$ & $  3.69 \pm  1.04$ & $  4.12 \pm  0.86$ & $  6.70 \pm  1.27$ & $  6.93 \pm  1.52$ & $  9.01 \pm  3.95$ & \\
& &Blue &15515& $  2.11 \pm  0.60$ & $  1.85 \pm  0.58$ & $  1.81 \pm  0.62$ & $  3.24 \pm  0.80$ & $  2.55 \pm  1.52$ & $  7.28 \pm  4.00$ & \\
\tableline

& &Red &4277& $  7.37 \pm  1.94$ & $  6.56 \pm  0.87$ & $  6.75 \pm  0.91$ & $  6.68 \pm  1.13$ & $  7.32 \pm  1.86$ & $ 15.07 \pm  5.09$ & \\
-19.5 & -20.5 &Green &1640& $  3.13 \pm  1.29$ & $  4.63 \pm  1.20$ & $  4.67 \pm  1.12$ & $  4.93 \pm  1.38$ & $  5.84 \pm  1.79$ & $ 11.06 \pm  3.81$ & \\
& &Blue &7669& $  4.88 \pm  0.92$ & $  3.18 \pm  0.74$ & $  2.56 \pm  0.53$ & $  2.96 \pm  0.94$ & $  4.88 \pm  1.60$ & $  9.98 \pm  5.30$ & \\
\tableline

& &Red &917& $ 13.45 \pm  3.28$ & $ 14.36 \pm  3.91$ & $ 11.63 \pm  2.80$ & $ 13.31 \pm  3.22$ & $ 14.47 \pm  3.20$ & $ 24.54 \pm  7.47$ & \\
-20.5 & -21.5 &Green &312& $ 10.10 \pm  4.58$ & $ 10.77 \pm  4.51$ & $  8.50 \pm  2.43$ & $  5.23 \pm  2.59$ & $  5.17 \pm  3.07$ & $  7.36 \pm  5.87$ & \\
& &Blue &1031& $  5.56 \pm  2.44$ & $  4.88 \pm  2.04$ & $  6.03 \pm  2.57$ & $  6.21 \pm  2.19$ & $  7.40 \pm  2.81$ & $ 20.49 \pm  6.96$ & \\
\tableline
 
\end{tabular}
\end{center}
\end{table*}

The luminosity dependent results for blue galaxies match the higher redshift DEEP2 results of \citet{Cooper06} and \citet{Coil08} more than the low redshift SDSS results, which show little luminosity dependence for blue galaxies (e.g., \citealt{Hogg03}). While not strictly a clustering study, \citet{Cooper06}, in their analysis of local environmental densities using a projected third-nearest-neighbor metric, found that blue galaxies showed as much of a luminosity trend as red galaxies at $z \sim 1$, and \citet{Coil08} found that blue galaxies exhibited luminosity-dependent clustering at small scales. Luminosity dependence in DEEP2 red galaxies was not conclusive due to larger error bars and a smaller effect size, but in at least one bin in projected distance ($r_{p} \sim  0.15$; see their Fig.~9), there is a hint of a similar trend. Additionally, \citet{Coil08} used four luminosity thresholds for each color subsample, but the brightest and faintest thresholds only differed by $\sim 1$ in median magnitude, and probe the $L^{*}$ regime, where luminosity dependence is weakest. Our results show that by exploring a wider range of luminosities, both blue and red galaxies show luminosity dependence down to $z \sim 0.5$. While Fig.~\ref{fig:fig6} appears to support a closer adherence to a power-law dependence on luminosity for blue galaxies compared red galaxies, further investigation of this is necessary to determine whether it is statistically significant. Regardless of the exact form of the dependence, luminosity dependence in the clustering of both red and blue galaxy populations should be considered a necessary ingredient in modeling galaxy formation. 

\subsection{Physics of Galaxy Evolution}

Both our color and luminosity dependent results have consequences for understanding the physics of galaxy assembly and evolution. We touch briefly on the most significant here.

In general agreement with work at both low and high redshift, we find a monotonic increase in clustering strength from the blue, through green, to the red galaxy samples (e.g., \citealt{Hogg03}; \citealt{Coil08}; \citealt{Zehavi11}). However, as mentioned in the previous subsection, we see signs that at the smallest scales, green galaxies are clustered similarly to blue galaxies, while at larger scales, they cluster as red galaxies. This could suggest suggest that they are not a distinct population, separate from blue and red galaxies, but that they are being quenched \citep{Krause13}. This clustering significance for the green galaxies at small scales is marginal, however, and as noted, the red and blue population are less distinct, and so physically, we cannot reliably distinguish it from the alternative for the green valley galaxies, which is that they are caused by scatter from the red and blue populations, or that they are dusty star-forming galaxies. However, the fact that the green galaxies have much more similar clustering to red than blue galaxies between $r = 250 \hkpc$ and $r = 1 \hmpc$ suggests that dust is an unlikely to explain most for the effect. Specifically, it appears that if the green valley is a mix of the red and blue populations, then the majority of the galaxies must be scattering in from the redder sample. On the other hand, if quenching is responsible for the signal, then satellites may be quenched as they enter the virial influence of a larger halo, while central galaxies may be only now be reaching a point at which mergers or internal processes are shutting off star formation. Given the parity in large-scale clustering with red galaxies, green galaxies are likely to be hosted by similar halos; a lower small-scale clustering, although very small in this work, agrees with the DEEP2 results of \citet{Coil08}, and the clustering signal of red spiral galaxies in \citet{Skibba09} (see their Fig.~10), which was smaller than for red galaxies generally at projected scales of $r_{p} < 1 \hmpc$.

We also find that luminosity dependence is strongest at the smallest scales, and weakest at scales of $r = 0.25 - 0.5 \hmpc$, for all our our color samples. However, the quantitatively similar fits to the luminosity dependence that we report in Fig.~\ref{fig:fig5} may belie different physical causes. In the two smallest bins in scale, red galaxies begin to show increased clustering (although not to the extent of \citealt{Zehavi11}) at faint luminosities, while to statistical significance, the blue galaxies' clustering amplitudes plateau. This should be unsurprising if external quenching mechanisms are at work: less luminous, blue galaxies will be correlated with star-forming galaxies with lower stellar mass and less total gas mass. {They will remain star-forming longer} if they live in matter underdense environments; if they are satellites in more matter overdense regions, physical quenching mechanisms will more quickly end star-formation, moving them onto the red sequence. Thus, more extant satellites will contribute to the red galaxy clustering signal, and the contributions to the blue signal will primarily be from the underdense field. While recent galactic conformity observations show that quenched fractions around passive hosts are higher than around star-forming hosts, these results are not in conflict. Physical quenching mechanisms will still move \textit{some} less-luminous blue satellites onto the red sequence, whether by mass-dependent quenching \citep{Phillips14} or via feedback \citep{Hartley15} or some other process. Thus, it is not necessary that all low-luminosity blue satellites are quenched; the shape of the clustering signal is expected if lower-luminosity blue galaxies quench more easily than more luminous blue galaxies.

Likewise, the relative bias of red and blue galaxies is particularly strong at $r = 0.25 \hmpc$, where there is an inflection in the clustering of the blue galaxy sample that is at least $\sim 2\sigma$ stronger than at the other scales. These scales roughly correspond to the transition between the one- and two-halo clustering regimes. \citet{Zheng07} fit HOD models to intermediate redshift DEEP2 observations and find the transition from the one- to two-halo term occurs at projected distances of $r_{p} \sim 0.4 - 0.6 \hmpc$, with the higher luminosity thresholds having larger transition distances. As those thresholds range from $M_B < -19$ to $M_B < -20.5$ (and noting that $M_g \sim M_B$), while the mean luminosities of the red and blue samples are $M_g = -19.10$ and $M_g = -18.67$, respectively, we would expect the halo transition to occur at a somewhat smaller physical scale. Thus, this signal may result from blue galaxies being found in less massive halos than their equal-luminosity, red counterparts. However, future modeling in mock halos of the cross-correlation statistic is clearly needed to confirm this as the likely cause.

\section{Conclusions}

We have measured the clustering of $62$,$106$ spectroscopic galaxies from the PRIMUS survey between $z = 0.2$ and $z = 1.0$. We used a method previously used principally for bright LRGs (E05; see also \citealt{Hogg03}) to measure the cross-correlation of roughly $L^{*}$ secondary galaxies around PRIMUS primary galaxies. We use this method to maximize the statistical power of our primary sample, while subdividing it as a function of color, luminosity, and scale. Our key results are that:

\begin{enumerate}
\item{We present precise and detailed measurements of strong --- but notably scale-dependent --- luminosity dependence in the clustering amplitude at scales of $r = 0.0625 - 2.0 \hmpc$. This luminosity dependence is evident separately for each of the red, green, and blue galaxy samples, showing that luminosity dependence in clustering is not a color-dependent phenomena, nor a side-effect of changing red fractions in higher luminosity samples.}
\item{Luminosity dependence is present over the full luminosity range for $r \le 0.125 \hmpc$, while at larger scales, the luminosity dependence is only evident at $M_g = -21$, while the clustering plateaus for fainter galaxies. Luminosity dependence is a thus a complicated emergent phenomena that traces the small-scale effects of galaxy formation and evolution. Reproducing color and luminosity binned clustering, in addition to average clustering, provides better constraints on the addition of new galaxy physics and feedback prescriptions in these models.}
\item{We find that red galaxies cluster more strongly than green galaxies and much more strongly than blue galaxies at an effective redshift of $z = 0.5$. On average, over physical scales of $r = 0.0625 - 2.0 \hmpc$, we detect a relative bias of red to blue galaxies of $\sim 2 - 3$ and a relative bias of red to green galaxies of $\sim 1.5$. This agrees with previous work at both lower and higher redshifts.}
\item{We detect a maximum relative bias of $3.8 \pm 0.9$ between red and blue galaxies at $r = 250 \hkpc$. We posit that the significantly non-power-law behavior of the blue galaxy correlation function at this scale that leads to this large relative bias is indicative of the one- to two-halo transition occurring at a smaller scales for blue galaxies.}
\end{enumerate}

This paper is the second in a series of papers quantifying the clustering properties of PRIMUS galaxies. It follows \citet{Skibba14}, which examined the color and luminosity dependence of the auto-correlation function, $w_{p}(r_p)$, out to $r_{p} = 30 \hmpc$. Additionally, \citet{MendezPREP} will measure the auto-correlations as a function of stellar mass and specific star formation rate. \citet{Skibba15} analyze the stellar mass dependent clustering of PRIMUS galaxies using analytic models and mock galaxy catalogs. \cite{BrayPREP} will use the real-space, cross-correlation statistics used in this work to probe three-dimensional galactic conformity. This follow-up paper (Paper II) will use both luminosity- and color-selected secondary galaxies to examine the red fractions around PRIMUS galaxies both in the `1-halo' and the `2-halo' regime.

\section*{Acknowledgments}

A.D.B. and D.J.E. acknowledge support from NSF grants AST-0607541 and 0908442. A.L.C. acknowledges support from NSF CAREER award AST-1055081. G.Z. acknowledges partial support provided by NASA through Hubble Fellowship grant HST-HF2-51351 awarded by the Space Telescope Science Institute, which is operated by the Association of Universities for Research in Astronomy, Inc., under contract NAS 5-26555. Additional PRIMUS funding has been provided by NSF grants AST-0607701, 0908246, and 0908354, and NASA grants 08-ADP08-0019, NNX09AC95G.  We thank the staff at Las Campanas Observatory, Chile, where data was taken with the 6.5~meter Magellan Telescopes, and we acknowledge Magellan telescope time provided by NOAO community access. We thank the CFHTLS, COSMOS, DEEP2 and SWIRE teams for their public data releases and/or access to early releases of their imaging catalogs.

\bibliographystyle{apj}

\end{document}